\documentclass[aps, prd, letterpaper, twocolumn, amsmath, amssymb, preprintnumbers, superscriptaddress, floatfix, nofootinbib, longbibliography]{revtex4-2}

\usepackage[T1]{fontenc}
\usepackage{graphicx}
\usepackage{amsmath}
\usepackage{amsfonts}
\usepackage{xurl} 
\usepackage[colorlinks=true, linkcolor=blue, urlcolor=blue, citecolor=blue, anchorcolor=blue]{hyperref}
\usepackage{bbold}
\usepackage{bbm}
\usepackage{orcidlink}
\usepackage{tikz-feynman}
\tikzfeynmanset{compat=1.0.0}

\newcommand{\ignore}[1]{}

\newcommand{\cL}{\ensuremath{\mathcal L} }

\newcommand{\Tr}[1]{\ensuremath{\mbox{Tr}\left[ #1 \right]} }

\newcommand{\beq}{\begin{equation}}
\newcommand{\eeq}{\end{equation}}
\newcommand{\beqs}{\begin{eqnarray}}
\newcommand{\eeqs}{\end{eqnarray}}

\newcommand{\orcidauthorINGOLDBY}{0000-0002-4690-3163} 
\newcommand{\orcidauthorPIAI}{0000-0002-2251-0111}

\begin{document}
\preprint{IPPP/25/89}

\title{Dilaton Effective Field Theory across the Conformal Edge}

\author{Thomas Appelquist\,
}
\email{thomas.appelquist@yale.edu}
\affiliation{Department of Physics, Sloane Laboratory, Yale University, New Haven, Connecticut 06520, USA}
\author{James Ingoldby\,\orcidlink{\orcidauthorINGOLDBY}}
\email{james.a.ingoldby@durham.ac.uk}
\affiliation{Institute for Particle Physics Phenomenology, Durham University, Durham DH1 3LE, UK}
\author{Maurizio Piai\,\orcidlink{\orcidauthorPIAI}}
\email{m.piai@swansea.ac.uk}
\affiliation{Department of Physics, Faculty  of Science and Engineering, Swansea University, Singleton Park, SA2 8PP, Swansea, United Kingdom}
\affiliation{Centre for Quantum Fields and Gravity, Faculty  of Science and Engineering, Swansea University, Singleton Park, SA2 8PP, Swansea, United Kingdom}



\begin{abstract}
Dilaton effective field theory (dEFT) can be employed to analyze lattice data in gauge theories that lie in close proximity of the lower edge of the conformal window. Under special conditions, we show that it can be used as a  diagnostic tool to distinguish near-conformal, yet confining,  theories from infrared conformal ones. We demonstrate this efficacy by analyzing two sets of lattice measurements taken from the literature. For the $SU(3)$ theory coupled to $N_f=8$ Dirac fermions transforming in the fundamental representation, our analysis favors confinement. For the $SU(2)$ theory with $N_f=1$ adjoint fermion, our fits favor infrared conformal behavior. We discuss future lattice measurements, and analysis refinements, that can further test this framework.

\end{abstract}

\maketitle

\section{Introduction}
\label{sec:intro}

In four dimensions, asymptotically free gauge theories are realized in different zero-temperature phases, depending on the type and number $N_f$ of light matter fields included. In the absence of matter fields, Yang-Mills theories confine, and the presence of few fermions does not alter qualitatively this feature, except for introducing continuous global symmetries and their spontaneous breaking. Conversely, for a large enough number of fermions, long distance physics is governed by the appearance of an infrared fixed point, with its associated conformal invariance. This  can be shown perturbatively, when $N_f$ is close to its upper bound insuring asymptotic freedom~\cite{Banks:1981nn}. The conformal behavior  is expected to persist over a finite range of $N_f>N_{f\,c}$, with the lower bound  defining the edge of the conformal window. The determination of $N_{f\,c}$, and the characterization of the theory near it, is an interesting strong-coupling problem.

Lattice field theory is the natural  instrument to study the problem, and in recent years it has led to significant advances---see the review in Ref.~\cite{Rummukainen:2022ekh} and references therein, as well as the more recent Refs.~\cite{Chung:2023mgr,Bennett:2023gbe,Nogradi:2023wnf,Bennett:2024tex,Athenodorou:2024rba}.
These studies explore a finite range of fermion masses that suppress infrared sensitivity to the lattice volume, and yield a discrete spectrum of bound states, as expected in confining theories. Some of these studies have reported evidence for the existence of a relatively light, flavor-singlet particle, together with a set of similarly light pseudoscalar particles, separated in mass from  heavier bound states~\cite{Appelquist:2007hu,Deuzeman:2008sc,Fodor:2009wk,LatKMI:2016xxi,Appelquist:2016viq,LatticeStrongDynamics:2018hun,Kotov:2021mgp,LatticeStrongDynamicsLSD:2021gmp,Hasenfratz:2022qan,LatticeStrongDynamics:2023bqp,LSD:2023uzj,LatKMI:2025kti,Fodor:2012ty,Fodor:2015vwa,Fodor:2016pls,Fodor:2017nlp,Fodor:2019vmw, Fodor:2020niv,Athenodorou:2014eua,Athenodorou:2016ndx,Athenodorou:2021wom,Athenodorou:2024rba}.

In a gauge theory outside of the conformal window, the pseudoscalars are naturally interpreted as pseudo-Nambu-Goldstone Bosons (pNGBs), associated with the spontaneous breaking of an approximate internal global symmetry acting on the matter fields. With the light flavor-singlet interpreted as a dilaton associated with the spontaneous breaking of  approximate dilatation symmetry, the behavior of a gauge theory taken to be just below the edge of the window has been successfully described employing dilaton effective field theory (dEFT)~\cite{Matsuzaki:2013eva,Golterman:2016lsd,Kasai:2016ifi,Hansen:2016fri,Golterman:2016cdd,Appelquist:2017wcg,Appelquist:2017vyy,Cata:2018wzl,Golterman:2018mfm,Cata:2019edh,Appelquist:2019lgk,Golterman:2020tdq,Golterman:2020utm}. See also the review in Ref.~\cite{Appelquist:2022mjb} and references therein, including the precursors in Refs.~\cite{Migdal:1982jp,Coleman:1985rnk}, the early study in the context of electroweak symmetry breaking~\cite{Goldberger:2007zk}, as well as recent phenomenological applications in Refs.~\cite{Appelquist:2020bqj,Appelquist:2022qgl,Cacciapaglia:2023kat,Appelquist:2024koa,Bruggisser:2022ofg,Alonso:2025ksv,Wu:2025hfp}, and theoretical developments in Refs.~\cite{Zwicky:2023krx,Faedo:2024zib,Elander:2025fpk,Cresswell-Hogg:2025kvr,Stegeman:2025sca,Stegeman:2025tdl}.

 In this paper, we expand on this idea, noting that the same dEFT can also be used just inside the conformal window, in the presence of a finite fermion mass. The dilaton and pNGBs continue to exist there as relatively light states. Therefore, dEFT can  be employed to explore near-edge gauge theories without an inside-versus-outside bias, and hence used as a diagnostic tool to discriminate between confining and conformal behavior. We test this framework by applying it to two gauge theories in proximity to the lower edge of the conformal window. We start with the well studied $SU(3)$ gauge theory with $N_f = 8$ Dirac fermions in the fundamental representation~\cite{LatticeStrongDynamics:2023bqp,LSD:2023uzj}, finding, as assumed in Refs.~\cite{Matsuzaki:2013eva,Golterman:2016lsd,Kasai:2016ifi,Hansen:2016fri,
Golterman:2016cdd,Appelquist:2017wcg,Appelquist:2017vyy,Cata:2018wzl,
Golterman:2018mfm,Cata:2019edh,Appelquist:2019lgk,Golterman:2020tdq,Golterman:2020utm}, that it is outside the conformal window. We then analyze the $SU(2)$ gauge  theory coupled to  $N_f =1$ Dirac fermion transforming in the adjoint representation~\cite{Athenodorou:2021wom,Athenodorou:2024rba}, finding evidence that it is inside. 

 In Section~\ref{sec:effectivetheory}, we discuss the general framework. We then implement it to fit the lattice data for the above two gauge theories in Section~\ref{sec:dfit}. In Section~\ref{sec:disc}, we summarize our results and discuss future directions.  We relegate to the Appendix some technical details about our fitting procedure, and tables reproducing the data we take from Refs.~\cite{LatticeStrongDynamics:2023bqp,SU(3)-data_release} and~\cite{Athenodorou:2021wom,SU(2)-1-data_release,SU(2)-2-data_release}.

\section{Dilaton Effective Field Theory}
\label{sec:effectivetheory}

In the dEFT Lagrangian density,  the canonically normalized dilaton field $\chi$  has a potential $V(\chi)$ and couplings to the pNGBs: 
\beqs
	\label{eq:L}
	\cL &=& \frac{1}{2}\partial_{\mu}\chi\partial^{\mu}\chi
    \, - \, V(\chi) 
    \, \\ 
    &&+ \frac{P}{4}\chi^2 \, \Tr{\partial_{\mu}\Sigma(\partial^{\mu}\Sigma)^{\dagger}} 
    + \frac{R}{ 4} \chi^y \, \Tr{\Sigma X^{\dagger}+  X\Sigma^{\dagger}} \, .
\nonumber
\eeqs
The matrix field $\Sigma$ describes the pNGBs. The factor $\chi^2$ multiplying their kinetic term acts as a conformal compensator, and $P$ is a positive dimensionless parameter to be fit to lattice data. The final term describes explicit breaking of the internal global symmetry arising from the finite mass $m$ assigned to the fermions in the underlying gauge theory. The $\chi$ dependence of the terms in Eq.~(\ref{eq:L}) follows from a spurion analysis, explained in Refs.~\cite{Appelquist:2019lgk,Appelquist:2022mjb}, where higher-order EFT corrections, including both loop effects and additional operators in the Lagrangian, are also discussed. Here we work at leading order in the EFT expansion, using the Lagrangian in Eq.~(\ref{eq:L}) at the tree level.

The form of $\Sigma$ and the parameter matrix $X$ are dictated by the internal symmetry and the appropriate breaking pattern. In the case of gauge theories with matter in complex representations (as with the $SU(3)$, $N_f=8$ theory), $\Sigma$ transforms on the left and right according to the $SU(N_{f})_L \times SU(N_{f})_R$ symmetry, and satisfies the nonlinear constraint $\Sigma \Sigma^{\dagger} = \mathbbm{1}_{N_f\times N_f}$,  while $X=\mathbbm{1}_{N_f\times N_f}$.  

For theories with  
matter in real representations, the global symmetry is enhanced to $SU(2N_f)$, and the pNGBs span the $SU(2N_f)/SO(2N_f)$
coset. Hence, $\Sigma\equiv S\,X=X\,S^T$ transforms as the  symmetric representation of $SU(2N_f)$, with $S$ satisfying $S S^{\dagger} = \mathbbm{1}_{2N_f\times 2N_f}$, and with $X$  symmetric and traceless. For the $SU(2)$ theory with adjoint matter ($N_f = 1$), we choose $X=\tau^1$, the first Pauli matrix.\footnote{For pseudo-real representations, $X$ and $\Sigma$ are antisymmetric matrices, and the pNGBs describe the $SU(2N_f)/Sp(2N_f)$ coset.}

The factor $\chi^y$ in the final term of $\cL$ is also a conformal compensator, with the scaling dimension $y$ determined by fits to lattice data. Near the edge of the conformal window, the anomalous dimension of the fermion bilinear $\bar{\psi}\psi$ is expected to approach 1~\cite{Cohen:1988sq}, implying $y\approx 2$. This value is also motivated in earlier dilaton EFT studies~\cite{Matsuzaki:2013eva}. The non-negative constant $R$, of dimension $4-y$, is induced by the presence of the underlying fermion mass $m$, defined at the lattice cutoff. It vanishes in the limit $m \rightarrow 0$, and we take its $m$-dependence to be linear: 
\begin{align}
 R = B_f m\,,
\end{align} 
where $B_f$ is a positive constant.

For the dilaton potential $V(\chi)$,  we adopt the form
\begin{align}
	V(\chi)  = A\chi^4 + B \chi^{\Delta}\,,
	\label{eq:vdelta}
\end{align}
where $\Delta$ ($>0$), $A$, and $B$ are real parameters, and where the second term describes the expected deformation of conformality present even in the limit $m = 0$. 
The full scalar potential is given by 
\begin{align}
	W(\chi) = V(\chi) - { \frac{1}{2}} x N_f R \chi^y\,,
    \label{eq:W}
\end{align}
with $x=1$ for complex representations, and $x=2$ for real and pseudo-real ones. For the case $y < 4, \Delta$, and with $R>0$,  the second term dominates for small $\chi$, driving the system away from the origin and toward a stable, symmetry breaking minimum of $W(\chi)$.

We note in passing that our previous studies~\cite{Appelquist:2019lgk,Appelquist:2022mjb} of the $SU(3)$ gauge theory with $N_f = 8$ fermions in the fundamental representation were conducted 
with a constraint on the parameters of the potential $V(\chi)$ such that it has a symmetry breaking minimum,
insuring that the spontaneous breaking of approximate conformal symmetry persists in the limit $R=0$. Here we proceed more generally, allowing lattice fits to determine whether or not $V(\chi)$ has a symmetry-breaking minimum, that is whether the theory is just outside or just inside the conformal window. Our approach allows for the possibility that $V(\chi)$ could approach a much-employed logarithmic form~\cite{Appelquist:2019lgk,Golterman:2020tdq,Golterman:2020utm,Appelquist:2022mjb,Migdal:1982jp,Coleman:1985rnk} in the marginal-deformation limit $\Delta\rightarrow4$.


\section{Fits to Lattice Data}
\label{sec:dfit}

The parameters of the dEFT in Eq.~(\ref{eq:L}) derive from the underlying gauge theory that UV completes it. They are directly related to the decay constants ($F_d$, $F_{\pi}$) and masses ($M_d$, $M_{\pi}$) of the dilaton and pNGBs. Lattice studies of the underlying gauge theory typically present values for these quantities as functions of a common mass $m$ of the fermions, defined at the UV cutoff.

The quantity $F_{d}$ is the vacuum expectation value (VEV) of $\chi$, determined in terms of the dEFT parameters by extremizing $W(\chi)$:
\begin{align}
	4 A F_{d}^{4-y} +\Delta B F_{d}^{\Delta-y} - { \frac{1}{2}}  x N_{f} R y= 0 \,.
\end{align}
The dilaton mass is given by 
\beqs
    M_d^2 &=& 12 A F_d^2 + \Delta(\Delta-1)B F_d^{\Delta - 2} \nonumber\\
    &&- { \frac{1}{2}} x N_{f} R\, y(y-1)  F_d^{y-2}\,.
\eeqs
With $\Sigma \equiv  \exp [2 i \Pi/F_{\pi}]$, canonical normalization of the pNGB kinetic term gives
\begin{align}
F_{\pi}^2 = PF_{d}^2\,,
\end{align} 
and
\begin{align}
M_{\pi}^2 = RF_{d}^{y} / F_{\pi}^2 \,.
\end{align} 

Lattice measurements exist for only the quantities $F_{\pi}^2$, $M_{\pi}^2$, and $M_d^2$, so we reorganize the above equations to obtain three useful fit equations. The first, independent of $V_{\Delta}(\chi)$, is 
\begin{align}
\label{eq1}
M_{\pi}^{2} F_{\pi}^{2-y} - C m = 0 \,.
\end{align} 
where $C \equiv B_f/P^{y/2}$ is treated as a fit parameter. Two additional equations can be obtained in which the exponential dependence on the fit parameter $y$ is removed. They take the form 
\begin{align}
\label{eq2}
\frac{M_{\pi}^{2}}{F_{\pi}^{2}} - \frac{2}{y x N_f P^2}\left[4A +\Delta B \left(\frac{F^2_\pi}{P}\right)^{\frac{\Delta-4}{2}}\right] = 0\,,
\end{align} 
and
\begin{align}
\label{eq3}
\frac{M_{d}^{2}}{F_{\pi}^{2}} - \frac{4(4-y)A}{P}-\frac{\Delta(\Delta-y)B}{P}\left(\frac{F_\pi^2}{P}\right)^{\frac{\Delta-4}{2}} = 0\,.
\end{align} 

As we employ Eqs.~(\ref{eq1})-(\ref{eq3}) to fit the lattice data, we will find it helpful to depict the form of $V^{\prime}(\chi)$ in a way that incorporates the uncertainty in the fit parameters. Noting that $V^{\prime}(F_d) = [ \frac{1}{2}  x y N_{f} P^{1/2}] ~ M_{\pi}^{2}F_{\pi}$, and employing Eq.~(\ref{eq2}), we can display the form of $V^{\prime}(F_{d})$, simply by plotting $M_{\pi}^{2}F_{\pi}$ versus $F_{\pi}$ ($ = P^{-1/2} F_d$). This plot can be extended  to show the form of $V^{\prime}(\chi)$ even at values of $\chi$ below a potentially non-vanishing VEV, where it is negative, by continuation of $F_{\pi}$ to small values.

\begin{figure}
    \centering
    \includegraphics[width=0.8\linewidth]{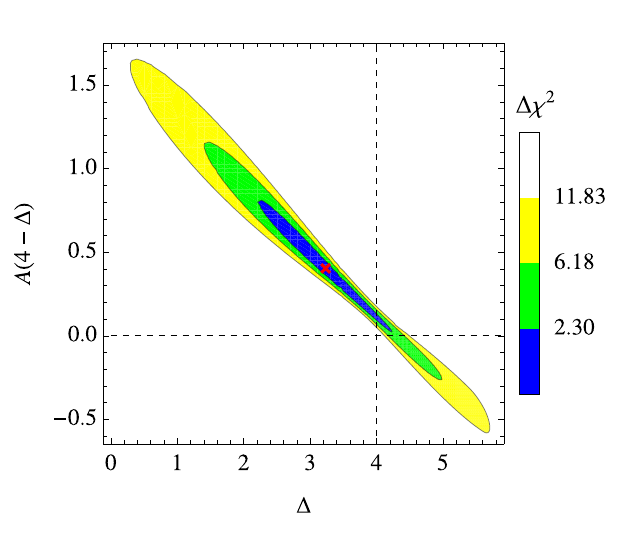}
    \caption {Contours of $\Delta\chi^2$ indicating confidence intervals (equivalent to $1
    \sigma$, $2\sigma$, and $3 \sigma$, respectively), for $A$ and $\Delta$ in the $SU(3)$ gauge theory with $N_f = 8$ Dirac fermions in the fundamental representation. The central value denoted by the red dot is reported in Table~\ref{Tab:deft}. For each point of the plot, we minimize the $\chi^2$ in respect to the other four fit parameters. The region $A > 0$ and $\Delta < 4$ is favored. }
    \label{fig:ADplot}
\end{figure}

\begin{figure}
    \centering
    \includegraphics[width=0.8\linewidth]{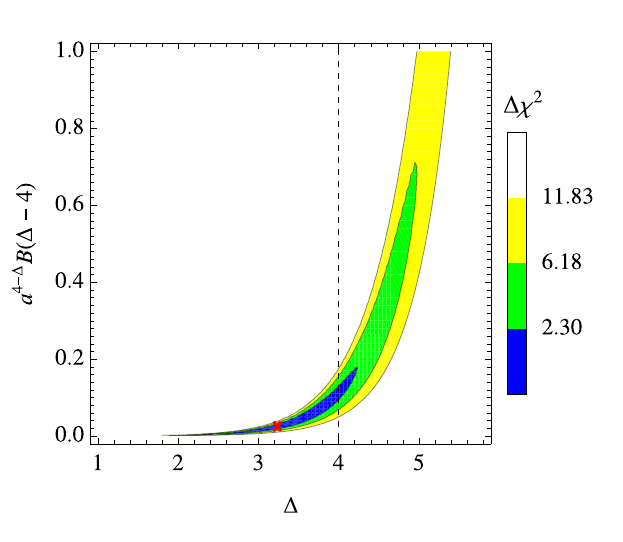}
    \caption{ Contours of $\Delta\chi^2$ indicating confidence intervals (equivalent to $1
    \sigma$, $2\sigma$, and $3 \sigma$, respectively), for $B$ and $\Delta$, in the $SU(3)$ gauge theory with $N_f = 8$ Dirac fermions in the fundamental representation. The central value denoted by the red dot is reported in Table~\ref{Tab:deft}. For each point of the plot, we minimize the $\chi^2$ in respect to the other four fit parameters. The region with $B < 0$ and $\Delta < 4$ is favored, consistent with being outside the conformal window.}
    \label{fig:BDplot}
\end{figure}

\begin{figure}
    \centering
        \includegraphics[width=0.8\linewidth]{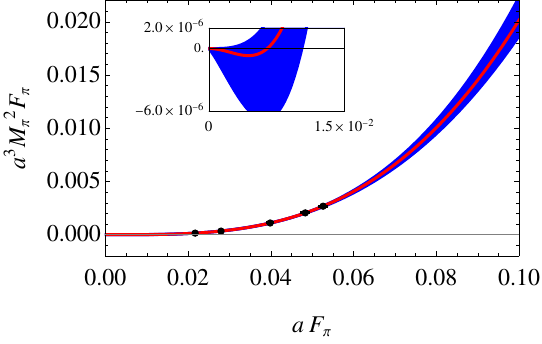}
        \caption{Plot of $a^3M^2_\pi F_\pi$ versus $a F_\pi$, in the $SU(3)$ theory with $N_f = 8$ fermions in the fundamental representation~\cite{LatticeStrongDynamics:2023bqp,SU(3)-data_release}. The red curve is based on the central values of Table~\ref{Tab:deft}, with the five data points denoted by the black discs. The blue shaded region represents the uncertainty in the fitting curve, equivalent to the  $1\sigma$ confidence level. The red curve reaches the horizontal axis at a nonzero value of $a F_{\pi}$, corresponding to the limit $m \rightarrow 0$ in a confining gauge theory. For larger values of $F_{\pi} \propto F_d$, we have $M_{\pi}^2 F_{\pi} \propto V^{\prime}(F_d)$. The curve is extrapolated downward, where it becomes negative, by continuing Eq.~(\ref{eq2}) to unphysical values of $F_{\pi}$, demonstrating that the region near the origin is unstable, as shown in the inset.}
    \label{fig:vshapenf8}
\end{figure}

\begin{table}
	\caption{Central values in the dEFT fit for the $SU(3)$ gauge theory with $N_f=8$ Dirac fermions in the fundamental representation. Lattice data for $M_{\pi}$, $M_d$, and $F_\pi$ is incorporated into this fit, for 5 different values of the underlying fermion mass $m$, corresponding to 15 data points. The lattice measurements  are taken from Refs.~\cite{LatticeStrongDynamics:2023bqp,SU(3)-data_release}, and reproduced in Table~\ref{Tab:3}. Dimensional quantities are presented in units of the lattice spacing, $a$. We do not quote individual uncertainties for all of the fit parameters here, as the likelihood in parameter space is non-Gaussian and exhibits strong correlations between parameters. Instead, joint confidence regions for $A$, $B$ and $\Delta$ are shown in Figs.~\ref{fig:ADplot} and \ref{fig:BDplot}. For $y$, we conservatively estimate the uncertainty to be $5\%$, reflecting the spread of acceptable fits.}
	\label{Tab:deft}
	\centering
	\vspace{12pt}
	\renewcommand\arraystretch{1.2}
	\begin{tabular}{| c | c |}
		\hline
		Parameter & Central Value \\
		\hline\hline
		$y$ & 2.081  \\
		$a^{3-y}C$ & 7.16 \\
		$\Delta$ & 3.237 \\
		$P$ & 0.1050 \\
		$A$ & 0.527 \\
		$a^{4-\Delta} B$ & - 0.0328\\
		\hline\hline
		$\chi^2/\text{dof}$ & 2.70 / 9 \\
		\hline
	\end{tabular}
\end{table}

\subsection{$SU(3)$ Gauge Theory with $N_f = 8$ Fermions in the Fundamental Representation}

The lattice data obtained from Refs.~\cite{LatticeStrongDynamics:2023bqp,SU(3)-data_release} is summarized in Table~\ref{Tab:3} of Appendix~\ref{app:data}. The masses and decay constants of the $63$ pNGBs and the masses of the dilaton are listed there for $5$ values of the underlying fermion mass $m$. As noted in the Appendix, 
correlations are neglected. The choice of lattice coupling $\beta$ and the range of fermion mass $m$ employed in the lattice study lead to masses $M_{\pi}$ and $M_d$ well below the inverse lattice spacing, and also below the masses of heavier states, which are not included in the dEFT analysis. 

Results for $A$, $B$, and $\Delta$ are shown in Figs.~\ref{fig:ADplot} and~\ref{fig:BDplot} in the form of $\Delta\chi^2$ contours. The favored upper left quadrants together correspond to values $\Delta < 4$, $A>0$, and $B<0$. The central values of all $6$ fit parameters are shown in Table~\ref{Tab:deft}. Those for $y$ and $P$ are consistent with earlier studies. The central values of $\Delta$ ($< 4)$, $A$ (positive), and $B$ (negative) imply that the theory lies just outside the conformal window, with $V(\chi)$ having a symmetry breaking minimum, as assumed in our earlier studies~\cite{Appelquist:2022mjb}. This is compatible with the expectation from the renormalization scheme invariant, multi-loop analysis of Ref.~\cite{Ryttov:2017kmx}, as well as the analysis of Ref.~\cite{Chung:2023mgr}, which combines lattice and perturbative inputs.

A graphical depiction of this conclusion is shown in Fig.~\ref{fig:vshapenf8}. We plot $M_{\pi}^{2}F_{\pi}$ as a function of $F_{\pi} \propto F_d$. The red curve is based on the central values of Table~\ref{Tab:deft}. It shows $M_{\pi}^{2}F_{\pi}$ vanishing at a finite value of $F_{\pi}$ (corresponding to $m \rightarrow 0$), and increasing continuously with increasing $F_{\pi}$. In this range, $M_{\pi}^{2}F_{\pi} \propto V^{\prime}(F_d)$.  We extend the red curve to lower values of $F_{\pi}$, where $V^{\prime} < 0$, by continuing the fit equations~(\ref{eq1})-(\ref{eq2}) into this regime. The blue shaded region represents uncertainty in the fitting curve equivalent to the $1 \sigma$ level. The five data points, depicted by the black discs, lie well above the value of $F_{\pi}$ where $V^{\prime}(F_d)$ vanishes, and well within a region where the uncertainty in the fit is small.

\subsection{$SU(2)$ Gauge Theory with $N_f = 1$ Fermion in the Adjoint Representation}

\begin{figure}[th]
    \centering
  \hspace{0.8cm}     \includegraphics[width=0.8\linewidth]{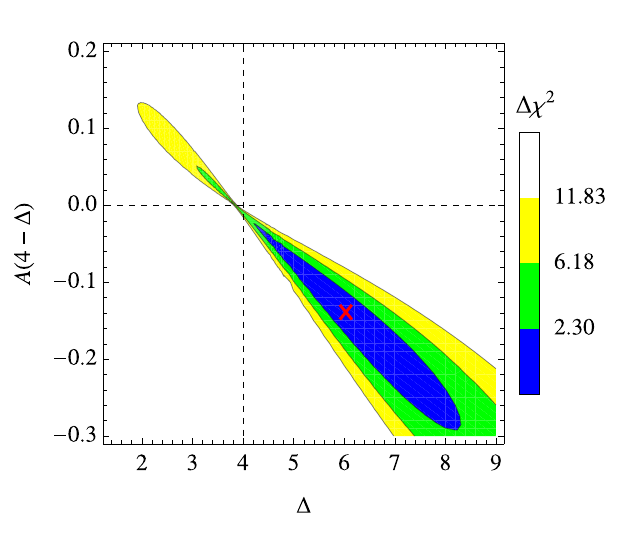}
    \caption{Contours of $\Delta\chi^2$ indicating confidence intervals (equivalent to $1
    \sigma$, $2\sigma$, and $3 \sigma$, respectively), for $A$ and $\Delta$ in the $SU(2)$ gauge theory with $N_f = 1$ Dirac fermion in the adjoint representation. The fit includes only data taken at the three lightest fermion mass points.  The central value is reported in Table~\ref{Tab:adjdeft}. For each point of the plot, we minimize the $\chi^2$ in respect to the other four fit parameters. The region with $\Delta>4$ and $A>0$ is favored, consistently with infrared conformal behavior.}
    \label{fig:AdAdj.pdf}
\end{figure}

\begin{figure}[h]
    \centering
   \hspace{0.8cm} \includegraphics[width=0.8\linewidth]{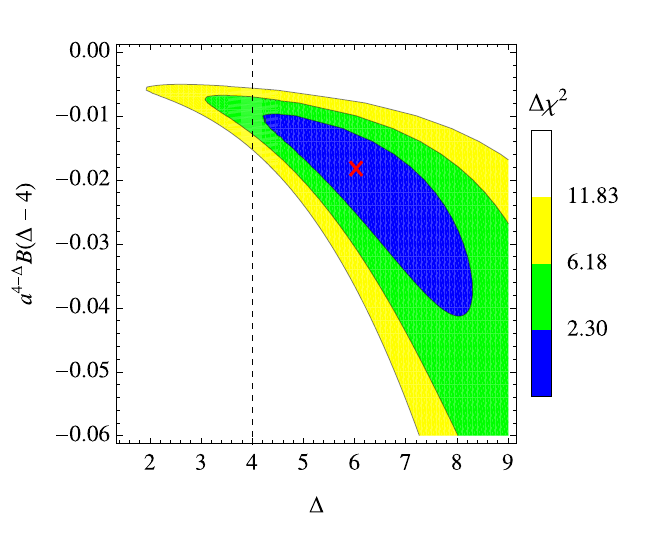}
    \caption{Contours of $\Delta\chi^2$ indicating confidence intervals (equivalent to $1
    \sigma$, $2\sigma$, and $3 \sigma$, respectively), for $B$ and $\Delta$ in the $SU(2)$ gauge theory with $N_f = 1$ Dirac fermion in the adjoint representation. The fit includes only data taken at the three lightest fermion mass points. The central value is reported in Table~\ref{Tab:adjdeft}. For each point of the plot, we minimize the $\chi^2$ in respect to the other four fit parameters. The region with $\Delta>4$ and $B<0$ is favored, consistently with infrared conformal behavior.}
    \label{fig:BdAdj.pdf}
\end{figure}

\begin{figure}
    \centering
    \includegraphics[width=0.8\linewidth]{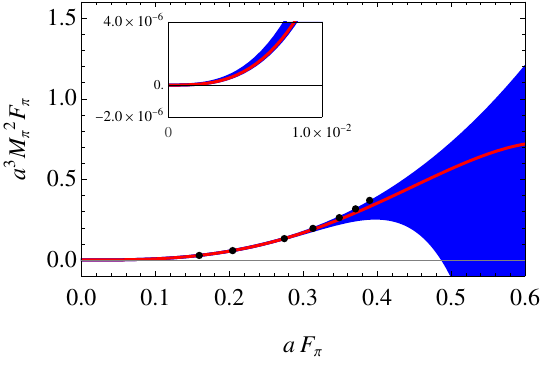}
      \caption{Plot of $a^3M^2_\pi F_\pi$ versus $a F_\pi \propto aF_d$, in the $SU(2)$ theory with $N_f = 1$ fermion in the adjoint representation ~\cite{Athenodorou:2021wom,SU(2)-1-data_release,SU(2)-2-data_release}, restricting attention to ensembles with $\beta=2.1$. The red curve is based on the central values of Table~\ref{Tab:adjdeft}, using only the three lightest ensembles. The data points are depicted by the black discs. The red curve reaches the horizontal line at $F_{\pi} = 0$, as shown in the inset, corresponding to the limit $m\rightarrow 0$ in an infrared conformal theory. Throughout the range, $a^3M^2_\pi F_\pi \propto V^{\prime}(F_d)$. The blue shaded region represents the uncertainty in the fitting curve, equivalent to the  $1\sigma$ confidence level.} 
      
    \label{fig:vshape}
\end{figure}

\begin{table}
	\caption{Central values for the dEFT fit for the $SU(2)$ gauge theory with $N_f=1$ Dirac fermion in the adjoint representation. The lattice data is taken from Refs.~\cite{Athenodorou:2021wom,SU(2)-1-data_release,SU(2)-2-data_release}, and reproduced in Table~\ref{Tab:2}, and has been restricted to the three ensembles with smallest fermion mass and $\beta = 2.1$. Lattice data for $M_{\pi}$, $M_d$, $F_\pi$, and $m_{\rm PCAC}$ has been incorporated into this fit, corresponding to 9 data points. Dimensional quantities are presented in units of the lattice spacing, $a$. We do not quote individual uncertainties for all the fit parameters due to the non-Gaussian, strongly correlated likelihood. Instead, joint confidence regions for $A$, $B$ and $\Delta$ are shown in Figs.~\ref{fig:AdAdj.pdf} and \ref{fig:BdAdj.pdf}, and we estimate an uncertainty of $10\%$ for $y$.}
	\label{Tab:adjdeft}
	\centering
	\vspace{12pt}
	\renewcommand\arraystretch{1.2}
	\begin{tabular}{| c |  c |}
		\hline
		Parameter & Central Value \\
		\hline\hline
		$y$ &   2.19 \\
		$a^{3-y}C$ &   7.69 \\
		$\Delta$ &  6.03\\
		$P$ &   0.131 \\
		$A$ &  0.0683\\
		$a^{4-\Delta} B$ &  -0.0090\\
		\hline\hline
		$\chi^2/\text{dof}$ &  0.616/3\\
		\hline
	\end{tabular}
\end{table}

We again employ Eqs.~(\ref{eq1})-(\ref{eq3}) to fit the lattice data~\cite{Athenodorou:2021wom,Athenodorou:2024rba,SU(2)-1-data_release,SU(2)-2-data_release}. In this case, with the spontaneous breaking pattern $SU(2) \rightarrow SO(2)$, there are $2$ pNGBs. 
We list the measurements performed with Wilson fermions and coupling $\beta=2.1$, for the largest available volumes,  in Table~\ref{Tab:2} of Appendix~\ref{app:data}. We identify the fermion mass $m_{\rm PCAC}$,  defined following Ref.~\cite{DelDebbio:2007wk}, with the mass parameter $m$ in our fits.  We include in our analysis only the three ensembles with the lightest mass, $a M_{\pi, d} < 0.7$. 
We tested the stability of the fits by including one or more heavier ensembles.\footnote{ As discussed in Ref.~\cite{Athenodorou:2024rba},
ensembles with $\beta>2.1$ yield a systematic decrease in the anomalous dimension of the fermion condensate,  indicating to us that 
in those ensembles the  Wilson fermion mass does not reach low enough values
 to assess the existence of an IR fixed point.
 }
 
Results for $A$, $B$, and $\Delta$ are shown in Figs.~\ref{fig:AdAdj.pdf} and \ref{fig:BdAdj.pdf} in the form of $\Delta \chi^2$ contours, with central values listed in Table~\ref{Tab:adjdeft}. Among the central values, those for $y$ and $P$ are similar to those of Table~\ref{Tab:deft}. 
In particular, $y \approx 2$,
  as expected near the edge of the conformal window~\cite{Cohen:1988sq}.
 We again have $A > 0$ and $B <0$, but now $\Delta > 4$. For these values, $B\chi^\Delta$ is an irrelevant operator that becomes less important at low energies. Although it explicitly breaks scale invariance, its effect on the full dilaton potential in Eq.~(\ref{eq:W}) diminishes in the massless limit, consistent with the claim that the gauge theory lies inside the conformal window. It is also the case that $V(\chi)$ turns down at large $\chi$, but the turn-down sets in well beyond the range of applicability of the dEFT. 

We depict this in Fig.~\ref{fig:vshape}, by plotting $M_{\pi}^2 F_{\pi} \propto V^{\prime}(F_d)$ as a function of $F_{\pi} \propto F_d$.  The red curve, based on the central values of Table~\ref{Tab:adjdeft}, shows $V^{\prime}(F_d)$ vanishing at $F_{\pi} = 0$ (corresponding to the limit $m \rightarrow 0$ in an infrared conformal gauge theory), and increasing with $F_{\pi}$ through values well beyond the location of the data points, depicted by the black discs. The data points lie low enough so that the $ 1 \sigma$-fit uncertainty, represented by the shaded region, is relatively small.

 We conclude that available lattice measurements indicate that this theory is inside the conformal window.
It is governed in the infrared by the extremum of $V(\chi)$ at $\chi = 0$, and all the masses and decay constants vanish in the limit $m=0$.   Additional lattice ensembles with small fermion mass and large $\beta$ will be helpful to confirm this preliminary conclusion.

\section{Summary and Discussion}
\label{sec:disc}

We employed dilaton effective field theory (dEFT) as a diagnostic tool to discriminate between infrared-conformal and near-conformal, confining gauge theories.
Our approach draws on spectroscopic lattice measurements performed with finite mass for the elementary fermions, in a range of parameter space in which both the pNGBs and the dilaton are clearly lighter than other states. This feature may be expected to emerge naturally in theories near the lower edge of the conformal window, for small enough fermion masses. We took the dEFT to apply smoothly through the transition between confinement and conformality.

We tested this approach with two lattice studies meeting these requirements. For the $SU(3)$ gauge theory coupled to $N_f=8$ Dirac fermions in the fundamental representation, we used measurements obtained with staggered fermions, restricted to the largest available lattice couplings and volumes~\cite{LatticeStrongDynamics:2023bqp,LSD:2023uzj}, to infer that the theory is outside the conformal window (near-conformal but confining in the $m \rightarrow 0$ limit). For the $SU(2)$ theory with $N_f=1$ fermions in the adjoint representation, we used data obtained with Wilson fermions~\cite{Athenodorou:2021wom,Athenodorou:2024rba}, by restricting attention  to ensembles in which  the relevant masses are significantly smaller than the lattice cutoff, to minimize lattice artifacts. Our analysis indicates that this theory is inside the conformal window (infrared conformal in the $m \rightarrow 0$ limit).

We made several simplifying assumptions. We ignored systematic effects in the lattice measurements, such as correlations between individual measurements, and contamination from finite-volume and finite size effects. We included only leading-order terms in the dEFT Lagrangian. 
We verified explicitly that the lattice measurements employed small enough fermion masses to explore only small-field dEFT excursions.

Our results should be taken as preliminary, as indicated by the large uncertainty in some of the outputs---in particular the scaling dimension $\Delta$. Future precision lattice measurements, particularly for the $SU(2)$ theory, could overcome these limitations, by exploring lattice parameters with smaller fermion masses and less contamination from lattice artifacts. 

The framework developed here nevertheless provides a systematic EFT description that can accommodate both confining an infrared-conformal possibilities on equal footing, and can be extended to incorporate additional observables as further input. It can also be applied to other gauge theories. It would be interesting, for example, to apply our analysis to the $SU(3)$ theory coupled to $N_f=10$ Dirac fermions in the fundamental representation~\cite{Kotov:2021mgp}, as well as to lattice data for the theory with $N_f=2$ sextet fermions~\cite{Fodor:2012ty,Fodor:2015vwa,Fodor:2016pls,Fodor:2017nlp,Fodor:2019vmw, Fodor:2020niv}, and to theories with other gauge groups, such as $Sp(4)$~\cite{Bennett:2023gbe,Bennett:2024tex}.

\begin{acknowledgments}

We thank Ed Bennett and Biagio Lucini for useful discussions. The work of  M.P. has been  supported in part by the STFC Consolidated Grants No.  ST/T000813/1 and ST/X000648/1. M.P. received funding from the European Research Council (ERC) under the European Union’s Horizon 2020 research and innovation program under Grant Agreement No.~813942. J.I. has been supported by STFC under grants No. ST/X000745/1 and No. ST/X003167/1.

\vspace{1.0cm}
{\bf Research Data Access Statement}---The 
data generated for this manuscript can be downloaded from Ref.~\cite{data_release}. 

\vspace{1.0cm}
{\bf Open Access Statement}---For the purpose of open access, the authors have applied a Creative Commons 
Attribution (CC BY) license to any Author Accepted Manuscript version arising.

\end{acknowledgments}

\begin{table}[t!]
	\caption{ Numerical lattice measurements taken from Table~IX of Ref.~\cite{LatticeStrongDynamics:2023bqp} with $\beta = 4.8$ (see also the data release~\cite{SU(3)-data_release}), for the $SU(3)$ theory coupled to $N_f=8$ fermions transforming in the fundamental representation. The errors in parenthesis includes  statistical and systematics, but correlations are neglected. The fermion mass,
	$a m$, appears as a parameter  in the lattice staggered action, and hence there is no associated
	  uncertainty. }
	\label{Tab:3}
	\centering
	\vspace{12pt}
	\renewcommand\arraystretch{1.2}
	\begin{tabular}{| c | c |c |c|}
		\hline
		$a m $ & $a M_{\pi}$ & $a F_{\pi}$ & $ a M_{d}$\\
		\hline\hline
        		0.00889 & 0.2253(37) &0.05262(88) & 0.301(12) \\
                		0.0075 & 0.2057(34) & 0.04831(80) & 0.2744(85) \\
                         		0.005 & 0.1657(27)  & 0.03982(66) & 0.2408(87) \\
                                		0.00222 & 0.1087(18) & 0.02794(47) & 0.1545(83) \\
		0.00125 & 0.0811(13) & 0.02168(36) & 0.1174(48) \\
		\hline
	\end{tabular}
\end{table}

\begin{table}[t!]
	\caption{ Numerical lattice measurements for the $SU(2)$ theory coupled to $N_f=1$ fermion transforming in the adjoint representation~\cite{Athenodorou:2021wom,SU(2)-1-data_release,SU(2)-2-data_release}, for
    ensembles  with $\beta=2.1$, and the largest available volumes. The errors in parenthesis are statistical only, and correlations are neglected. These numerical results are obtained with Wilson fermions, hence a measure of symmetry breaking effects due to finite fermion mass is provided by the PCAC mass, $a m_{\rm PCAC}$, extracted according to the process described in Ref.~\cite{DelDebbio:2007wk}.  }
	\label{Tab:2}
	\centering
	\vspace{12pt}
	\renewcommand\arraystretch{1.2}
	\begin{tabular}{| c | c |c |c|}
		\hline
		$a m_{\rm PCAC}$ & $a M_{\pi}$ & $a F_{\pi}$ & $ a M_{d}$\\
		\hline\hline
		0.15326(21) & 0.97363(51)& 0.38982(94)& 0.847(34)\\
 0.13820(17) &  0.92461(45)&  0.37075(71)& 0.716(57)\\
 0.12200(23) &  0.86745(64)&  0.34875(92)& 0.690(38)\\
 0.10251(16) &  0.79101(51)&  0.31313(97)& 0.565(36)\\		
  0.08011(19) & 0.69504(63) & 0.27437(70) & 0.469(33) \\
  0.04969(12) & 0.53319(48) & 0.20450(67) & 0.373(29) \\
  0.03246(14) &0.42018(32) & 0.15925(36) & 0.295(13)\\
		\hline
	\end{tabular}
\end{table}

\appendix
\section{Maximum likelihood analysis and lattice data}
\label{app:data}

We here collect some technical details needed to reproduce our main results.
First, we summarize in Tables~\ref{Tab:3} and~\ref{Tab:2} the spectroscopic measurements for  pNGBs and  dilaton, 
taken from Refs.~\cite{LatticeStrongDynamics:2023bqp} and~\cite{Athenodorou:2021wom}.
These measurements are combined with Eqs.~(\ref{eq1}), (\ref{eq2}), and~(\ref{eq3}), to define the $\chi^2$, in the following way. 

Given the $N$ fitting functions, $\overrightarrow{f}=\{f_i\}_{i=1}^{N}$, which depend on the $K$ parameters of the model, $\overrightarrow{a}=\{a_k\}_{k=1}^{K}$, and on $M$ experimentally  measurable 
quantities, $\overrightarrow{x}=\{x_j\}_{j=1}^{M}$, we define the $N\times N$
covariance matrix $\bar{\cal C}$ as
\beqs
\bar{\cal C}&\equiv&O^T\,{\cal C}\, O\,,
\eeqs
where ${\cal C}$ is the $M\times M$ covariance matrix of the observables, and $O_{j}^{\,\,i}=\frac{\partial f_i}{\partial x_j}$ is
the $M\times N$ matrix of partial derivatives. The 
$\chi^2$ function is the  following
\beqs
\chi^2(\overrightarrow{a})&\equiv&\overrightarrow{f}^T(\overrightarrow{a},\overrightarrow{x})\, {\bar{\cal C}}^{-1}\,\overrightarrow{f}(\overrightarrow{a},\overrightarrow{x})\,,
\eeqs
The best fit value for $\overrightarrow{a}$ is  obtained by minimizing numerically the function $\chi^2(\overrightarrow{a})$.

As we do not have access to the covariance matrix for the measurements listed in Tables~\ref{Tab:3} and~\ref{Tab:2}, we take ${\cal C}$ to be diagonal, with non-vanishing elements given by the variance due to the statistical errors of the measurements, $\sigma^2_j$. We do not include systematic effects, assuming that they are smaller than the statistical ones, for those measurements we retain in the analysis.  The functions $f_i$ are defined by the left-hand side of Eqs.~(\ref{eq1}), (\ref{eq2}), and~(\ref{eq3}), respectively. For the $SU(3)$ theory, we treat the $5$ measurements of $a M_{\pi}$, $aF_{\pi}$, and $aM_d$ as the observables. The five choices of $a m $ are treated as numerical values of a parameter that is known without uncertainty --- the mass of the staggered fermions in the lattice action. For the $SU(2)$ theory, our $\chi^2$ analysis makes use of the measurements of $am_{\rm PCAC}$, $aM_{\pi}$, $a F_{\pi}$, and $aM_d$, restricting attention to the three ensembles with the lightest masses, among those with $\beta=2.1$.

\bibliographystyle{JHEP_custom} 
\bibliography{ref}

\providecommand{\href}[2]{#2}\begingroup\raggedright\begin{thebibliography}{10}

\bibitem{Banks:1981nn}
T.~Banks and A.~Zaks, \emph{{On the Phase Structure of Vector-Like Gauge
  Theories with Massless Fermions}},
  \href{https://doi.org/10.1016/0550-3213(82)90035-9}{\emph{Nucl. Phys. B}
  {\bfseries 196} (1982) 189}.

\bibitem{Rummukainen:2022ekh}
K.~Rummukainen and K.~Tuominen, \emph{{Lattice Computations for Beyond Standard
  Model Physics}},
  \href{https://doi.org/10.3390/universe8030188}{\emph{Universe} {\bfseries 8}
  (2022) 188}.

\bibitem{Chung:2023mgr}
H.S.~Chung and D.~Nogradi,
  \emph{{f{\ensuremath{\varrho}}/m{\ensuremath{\varrho}} and
  f{\ensuremath{\pi}}/m{\ensuremath{\varrho}} ratios and the conformal
  window}}, \href{https://doi.org/10.1103/PhysRevD.107.074039}{\emph{Phys. Rev.
  D} {\bfseries 107} (2023) 074039}
  [\href{https://arxiv.org/abs/2302.06411}{{\ttfamily 2302.06411}}].

\bibitem{Bennett:2023gbe}
E.~Bennett et~al., \emph{{Symplectic lattice gauge theories in the grid
  framework: Approaching the conformal window}},
  \href{https://doi.org/10.1103/PhysRevD.108.094508}{\emph{Phys. Rev. D}
  {\bfseries 108} (2023) 094508}
  [\href{https://arxiv.org/abs/2306.11649}{{\ttfamily 2306.11649}}].

\bibitem{Nogradi:2023wnf}
D.~Nogradi and H.S.~Chung, \emph{{Mesonic decay constant and mass ratios and
  the conformal window}}, \href{https://doi.org/10.22323/1.453.0084}{\emph{PoS}
  {\bfseries LATTICE2023} (2024) 084}
  [\href{https://arxiv.org/abs/2311.08668}{{\ttfamily 2311.08668}}].

\bibitem{Bennett:2024tex}
E.~Bennett, D.K.~Hong, H.~Hsiao, J.-W.~Lee, C.J.D.~Lin, B.~Lucini et~al.,
  \emph{{Meson spectroscopy in the Sp(4) gauge theory with three antisymmetric
  fermions}}, \href{https://doi.org/10.1103/PhysRevD.111.074511}{\emph{Phys.
  Rev. D} {\bfseries 111} (2025) 074511}
  [\href{https://arxiv.org/abs/2412.01170}{{\ttfamily 2412.01170}}].

\bibitem{Athenodorou:2024rba}
A.~Athenodorou, E.~Bennett, G.~Bergner, P.~Butti, J.~Lenz and B.~Lucini,
  \emph{{SU(2) gauge theory with one and two adjoint fermions towards the
  continuum limit}},  \href{https://arxiv.org/abs/2408.00171}{{\ttfamily
  2408.00171}}.

\bibitem{Appelquist:2007hu}
T.~Appelquist, G.T.~Fleming and E.T.~Neil, \emph{{Lattice study of the
  conformal window in QCD-like theories}},
  \href{https://doi.org/10.1103/PhysRevLett.100.171607}{\emph{Phys. Rev. Lett.}
  {\bfseries 100} (2008) 171607}
  [\href{https://arxiv.org/abs/0712.0609}{{\ttfamily 0712.0609}}].

\bibitem{Deuzeman:2008sc}
A.~Deuzeman, M.P.~Lombardo and E.~Pallante, \emph{{The Physics of eight
  flavours}}, \href{https://doi.org/10.1016/j.physletb.2008.10.039}{\emph{Phys.
  Lett. B} {\bfseries 670} (2008) 41}
  [\href{https://arxiv.org/abs/0804.2905}{{\ttfamily 0804.2905}}].

\bibitem{Fodor:2009wk}
Z.~Fodor, K.~Holland, J.~Kuti, D.~Nogradi and C.~Schroeder, \emph{{Nearly
  conformal gauge theories in finite volume}},
  \href{https://doi.org/10.1016/j.physletb.2009.10.040}{\emph{Phys. Lett. B}
  {\bfseries 681} (2009) 353}
  [\href{https://arxiv.org/abs/0907.4562}{{\ttfamily 0907.4562}}].

\bibitem{LatKMI:2016xxi}
{\scshape LatKMI} collaboration, \emph{{Light flavor-singlet scalars and
  walking signals in $N_f=8$ QCD on the lattice}},
  \href{https://doi.org/10.1103/PhysRevD.96.014508}{\emph{Phys. Rev. D}
  {\bfseries 96} (2017) 014508}
  [\href{https://arxiv.org/abs/1610.07011}{{\ttfamily 1610.07011}}].

\bibitem{Appelquist:2016viq}
T.~Appelquist et~al., \emph{{Strongly interacting dynamics and the search for
  new physics at the LHC}},
  \href{https://doi.org/10.1103/PhysRevD.93.114514}{\emph{Phys. Rev. D}
  {\bfseries 93} (2016) 114514}
  [\href{https://arxiv.org/abs/1601.04027}{{\ttfamily 1601.04027}}].

\bibitem{LatticeStrongDynamics:2018hun}
{\scshape Lattice Strong Dynamics} collaboration, \emph{{Nonperturbative
  investigations of SU(3) gauge theory with eight dynamical flavors}},
  \href{https://doi.org/10.1103/PhysRevD.99.014509}{\emph{Phys. Rev. D}
  {\bfseries 99} (2019) 014509}
  [\href{https://arxiv.org/abs/1807.08411}{{\ttfamily 1807.08411}}].

\bibitem{Kotov:2021mgp}
A.Y.~Kotov, D.~Nogradi, K.K.~Szabo and L.~Szikszai, \emph{{More on the flavor
  dependence of m$_{\rho}$/f$_{\pi}$}},
  \href{https://doi.org/10.1007/JHEP07(2021)202}{\emph{JHEP} {\bfseries 07}
  (2021) 202} [\href{https://arxiv.org/abs/2107.05996}{{\ttfamily
  2107.05996}}].

\bibitem{LatticeStrongDynamicsLSD:2021gmp}
{\scshape Lattice Strong Dynamics (LSD)} collaboration, \emph{{Goldstone boson
  scattering with a light composite scalar}},
  \href{https://doi.org/10.1103/PhysRevD.105.034505}{\emph{Phys. Rev. D}
  {\bfseries 105} (2022) 034505}
  [\href{https://arxiv.org/abs/2106.13534}{{\ttfamily 2106.13534}}].

\bibitem{Hasenfratz:2022qan}
A.~Hasenfratz, \emph{{Emergent strongly coupled ultraviolet fixed point in four
  dimensions with eight K\"ahler-Dirac fermions}},
  \href{https://doi.org/10.1103/PhysRevD.106.014513}{\emph{Phys. Rev. D}
  {\bfseries 106} (2022) 014513}
  [\href{https://arxiv.org/abs/2204.04801}{{\ttfamily 2204.04801}}].

\bibitem{LatticeStrongDynamics:2023bqp}
{\scshape Lattice Strong Dynamics} collaboration, \emph{{Light scalar meson and
  decay constant in SU(3) gauge theory with eight dynamical flavors}},
  \href{https://doi.org/10.1103/PhysRevD.110.054501}{\emph{Phys. Rev. D}
  {\bfseries 110} (2024) 054501}
  [\href{https://arxiv.org/abs/2306.06095}{{\ttfamily 2306.06095}}].

\bibitem{LSD:2023uzj}
{\scshape LSD} collaboration, \emph{{Hidden conformal symmetry from the
  lattice}}, \href{https://doi.org/10.1103/PhysRevD.108.L091505}{\emph{Phys.
  Rev. D} {\bfseries 108} (2023) L091505}
  [\href{https://arxiv.org/abs/2305.03665}{{\ttfamily 2305.03665}}].

\bibitem{LatKMI:2025kti}
{\scshape LatKMI} collaboration, \emph{{Novel view of the flavor-singlet
  spectrum from multi-flavor QCD on the lattice}},
  \href{https://doi.org/10.1103/vnml-g6nx}{\emph{Phys. Rev. D} {\bfseries 112}
  (2025) 114503} [\href{https://arxiv.org/abs/2505.08658}{{\ttfamily
  2505.08658}}].

\bibitem{Fodor:2012ty}
Z.~Fodor, K.~Holland, J.~Kuti, D.~Nogradi, C.~Schroeder and C.H.~Wong,
  \emph{{Can the nearly conformal sextet gauge model hide the Higgs
  impostor?}},
  \href{https://doi.org/10.1016/j.physletb.2012.10.079}{\emph{Phys. Lett. B}
  {\bfseries 718} (2012) 657}
  [\href{https://arxiv.org/abs/1209.0391}{{\ttfamily 1209.0391}}].

\bibitem{Fodor:2015vwa}
Z.~Fodor, K.~Holland, J.~Kuti, S.~Mondal, D.~Nogradi and C.H.~Wong,
  \emph{{Toward the minimal realization of a light composite Higgs}},
  \href{https://doi.org/10.22323/1.214.0244}{\emph{PoS} {\bfseries LATTICE2014}
  (2015) 244} [\href{https://arxiv.org/abs/1502.00028}{{\ttfamily
  1502.00028}}].

\bibitem{Fodor:2016pls}
Z.~Fodor, K.~Holland, J.~Kuti, S.~Mondal, D.~Nogradi and C.H.~Wong,
  \emph{{Status of a minimal composite Higgs theory}},
  \href{https://doi.org/10.22323/1.251.0219}{\emph{PoS} {\bfseries LATTICE2015}
  (2016) 219} [\href{https://arxiv.org/abs/1605.08750}{{\ttfamily
  1605.08750}}].

\bibitem{Fodor:2017nlp}
Z.~Fodor, K.~Holland, J.~Kuti, D.~Nogradi and C.H.~Wong, \emph{{The
  twelve-flavor $\beta$-function and dilaton tests of the sextet scalar}},
  \href{https://doi.org/10.1051/epjconf/201817508015}{\emph{EPJ Web Conf.}
  {\bfseries 175} (2018) 08015}
  [\href{https://arxiv.org/abs/1712.08594}{{\ttfamily 1712.08594}}].

\bibitem{Fodor:2019vmw}
Z.~Fodor, K.~Holland, J.~Kuti and C.H.~Wong, \emph{{Tantalizing dilaton tests
  from a near-conformal EFT}},
  \href{https://doi.org/10.22323/1.334.0196}{\emph{PoS} {\bfseries LATTICE2018}
  (2019) 196} [\href{https://arxiv.org/abs/1901.06324}{{\ttfamily
  1901.06324}}].

\bibitem{Fodor:2020niv}
Z.~Fodor, K.~Holland, J.~Kuti and C.H.~Wong, \emph{{Dilaton EFT from p-regime
  to RMT in the $\epsilon$-regime}},
  \href{https://doi.org/10.22323/1.363.0246}{\emph{PoS} {\bfseries LATTICE2019}
  (2020) 246} [\href{https://arxiv.org/abs/2002.05163}{{\ttfamily
  2002.05163}}].

\bibitem{Athenodorou:2014eua}
A.~Athenodorou, E.~Bennett, G.~Bergner and B.~Lucini, \emph{{Infrared regime of
  SU(2) with one adjoint Dirac flavor}},
  \href{https://doi.org/10.1103/PhysRevD.91.114508}{\emph{Phys. Rev. D}
  {\bfseries 91} (2015) 114508}
  [\href{https://arxiv.org/abs/1412.5994}{{\ttfamily 1412.5994}}].

\bibitem{Athenodorou:2016ndx}
A.~Athenodorou, E.~Bennett, G.~Bergner, D.~Elander, C.J.D.~Lin, B.~Lucini
  et~al., \emph{{Large mass hierarchies from strongly-coupled dynamics}},
  \href{https://doi.org/10.1007/JHEP06(2016)114}{\emph{JHEP} {\bfseries 06}
  (2016) 114} [\href{https://arxiv.org/abs/1605.04258}{{\ttfamily
  1605.04258}}].

\bibitem{Athenodorou:2021wom}
A.~Athenodorou, E.~Bennett, G.~Bergner and B.~Lucini, \emph{{Investigating the
  conformal behavior of SU(2) with one adjoint Dirac flavor}},
  \href{https://doi.org/10.1103/PhysRevD.104.074519}{\emph{Phys. Rev. D}
  {\bfseries 104} (2021) 074519}
  [\href{https://arxiv.org/abs/2103.10485}{{\ttfamily 2103.10485}}].

\bibitem{Matsuzaki:2013eva}
S.~Matsuzaki and K.~Yamawaki, \emph{{Dilaton Chiral Perturbation Theory:
  Determining the Mass and Decay Constant of the Technidilaton on the
  Lattice}}, \href{https://doi.org/10.1103/PhysRevLett.113.082002}{\emph{Phys.
  Rev. Lett.} {\bfseries 113} (2014) 082002}
  [\href{https://arxiv.org/abs/1311.3784}{{\ttfamily 1311.3784}}].

\bibitem{Golterman:2016lsd}
M.~Golterman and Y.~Shamir, \emph{{Low-energy effective action for pions and a
  dilatonic meson}},
  \href{https://doi.org/10.1103/PhysRevD.94.054502}{\emph{Phys. Rev. D}
  {\bfseries 94} (2016) 054502}
  [\href{https://arxiv.org/abs/1603.04575}{{\ttfamily 1603.04575}}].

\bibitem{Kasai:2016ifi}
A.~Kasai, K.-i.~Okumura and H.~Suzuki, \emph{{A dilaton-pion mass relation}},
  \href{https://arxiv.org/abs/1609.02264}{{\ttfamily 1609.02264}}.

\bibitem{Hansen:2016fri}
M.~Hansen, K.~Lang\ae{}ble and F.~Sannino, \emph{{Extending Chiral Perturbation
  Theory with an Isosinglet Scalar}},
  \href{https://doi.org/10.1103/PhysRevD.95.036005}{\emph{Phys. Rev. D}
  {\bfseries 95} (2017) 036005}
  [\href{https://arxiv.org/abs/1610.02904}{{\ttfamily 1610.02904}}].

\bibitem{Golterman:2016cdd}
M.~Golterman and Y.~Shamir, \emph{{Effective pion mass term and the trace
  anomaly}}, \href{https://doi.org/10.1103/PhysRevD.95.016003}{\emph{Phys. Rev.
  D} {\bfseries 95} (2017) 016003}
  [\href{https://arxiv.org/abs/1611.04275}{{\ttfamily 1611.04275}}].

\bibitem{Appelquist:2017wcg}
T.~Appelquist, J.~Ingoldby and M.~Piai, \emph{{Dilaton EFT Framework For
  Lattice Data}}, \href{https://doi.org/10.1007/JHEP07(2017)035}{\emph{JHEP}
  {\bfseries 07} (2017) 035}
  [\href{https://arxiv.org/abs/1702.04410}{{\ttfamily 1702.04410}}].

\bibitem{Appelquist:2017vyy}
T.~Appelquist, J.~Ingoldby and M.~Piai, \emph{{Analysis of a Dilaton EFT for
  Lattice Data}}, \href{https://doi.org/10.1007/JHEP03(2018)039}{\emph{JHEP}
  {\bfseries 03} (2018) 039}
  [\href{https://arxiv.org/abs/1711.00067}{{\ttfamily 1711.00067}}].

\bibitem{Cata:2018wzl}
O.~Cat\`a, R.J.~Crewther and L.C.~Tunstall, \emph{{Crawling technicolor}},
  \href{https://doi.org/10.1103/PhysRevD.100.095007}{\emph{Phys. Rev. D}
  {\bfseries 100} (2019) 095007}
  [\href{https://arxiv.org/abs/1803.08513}{{\ttfamily 1803.08513}}].

\bibitem{Golterman:2018mfm}
M.~Golterman and Y.~Shamir, \emph{{Large-mass regime of the dilaton-pion
  low-energy effective theory}},
  \href{https://doi.org/10.1103/PhysRevD.98.056025}{\emph{Phys. Rev. D}
  {\bfseries 98} (2018) 056025}
  [\href{https://arxiv.org/abs/1805.00198}{{\ttfamily 1805.00198}}].

\bibitem{Cata:2019edh}
O.~Cat\`a and C.~M\"uller, \emph{{Chiral effective theories with a light scalar
  at one loop}},
  \href{https://doi.org/10.1016/j.nuclphysb.2020.114938}{\emph{Nucl. Phys. B}
  {\bfseries 952} (2020) 114938}
  [\href{https://arxiv.org/abs/1906.01879}{{\ttfamily 1906.01879}}].

\bibitem{Appelquist:2019lgk}
T.~Appelquist, J.~Ingoldby and M.~Piai, \emph{{Dilaton potential and lattice
  data}}, \href{https://doi.org/10.1103/PhysRevD.101.075025}{\emph{Phys. Rev.
  D} {\bfseries 101} (2020) 075025}
  [\href{https://arxiv.org/abs/1908.00895}{{\ttfamily 1908.00895}}].

\bibitem{Golterman:2020tdq}
M.~Golterman, E.T.~Neil and Y.~Shamir, \emph{{Application of dilaton chiral
  perturbation theory to $N_f=8$, ${\rm SU}(3)$ spectral data}},
  \href{https://doi.org/10.1103/PhysRevD.102.034515}{\emph{Phys. Rev. D}
  {\bfseries 102} (2020) 034515}
  [\href{https://arxiv.org/abs/2003.00114}{{\ttfamily 2003.00114}}].

\bibitem{Golterman:2020utm}
M.~Golterman and Y.~Shamir, \emph{{Explorations beyond dilaton chiral
  perturbation theory in the eight-flavor SU(3) gauge theory}},
  \href{https://doi.org/10.1103/PhysRevD.102.114507}{\emph{Phys. Rev. D}
  {\bfseries 102} (2020) 114507}
  [\href{https://arxiv.org/abs/2009.13846}{{\ttfamily 2009.13846}}].

\bibitem{Appelquist:2022mjb}
T.~Appelquist, J.~Ingoldby and M.~Piai, \emph{{Dilaton Effective Field
  Theory}}, \href{https://doi.org/10.3390/universe9010010}{\emph{Universe}
  {\bfseries 9} (2023) 10} [\href{https://arxiv.org/abs/2209.14867}{{\ttfamily
  2209.14867}}].

\bibitem{Migdal:1982jp}
A.A.~Migdal and M.A.~Shifman, \emph{{Dilaton Effective Lagrangian in
  Gluodynamics}},
  \href{https://doi.org/10.1016/0370-2693(82)90089-2}{\emph{Phys. Lett. B}
  {\bfseries 114} (1982) 445}.

\bibitem{Coleman:1985rnk}
S.~Coleman, \emph{{Aspects of Symmetry}: {Selected Erice Lectures}}, Cambridge
  University Press, Cambridge, U.K. (1985),
  \href{https://doi.org/10.1017/CBO9780511565045}{10.1017/CBO9780511565045}.

\bibitem{Goldberger:2007zk}
W.D.~Goldberger, B.~Grinstein and W.~Skiba, \emph{{Distinguishing the Higgs
  boson from the dilaton at the Large Hadron Collider}},
  \href{https://doi.org/10.1103/PhysRevLett.100.111802}{\emph{Phys. Rev. Lett.}
  {\bfseries 100} (2008) 111802}
  [\href{https://arxiv.org/abs/0708.1463}{{\ttfamily 0708.1463}}].

\bibitem{Appelquist:2020bqj}
T.~Appelquist, J.~Ingoldby and M.~Piai, \emph{{Nearly Conformal Composite Higgs
  Model}}, \href{https://doi.org/10.1103/PhysRevLett.126.191804}{\emph{Phys.
  Rev. Lett.} {\bfseries 126} (2021) 191804}
  [\href{https://arxiv.org/abs/2012.09698}{{\ttfamily 2012.09698}}].

\bibitem{Appelquist:2022qgl}
T.~Appelquist, J.~Ingoldby and M.~Piai, \emph{{Composite two-Higgs doublet
  model from dilaton effective field theory}},
  \href{https://doi.org/10.1016/j.nuclphysb.2022.115930}{\emph{Nucl. Phys. B}
  {\bfseries 983} (2022) 115930}
  [\href{https://arxiv.org/abs/2205.03320}{{\ttfamily 2205.03320}}].

\bibitem{Cacciapaglia:2023kat}
G.~Cacciapaglia, D.Y.~Cheong, A.~Deandrea, W.~Isnard and S.C.~Park,
  \emph{{Composite hybrid inflation: dilaton and waterfall pions}},
  \href{https://doi.org/10.1088/1475-7516/2023/10/063}{\emph{JCAP} {\bfseries
  10} (2023) 063} [\href{https://arxiv.org/abs/2307.01852}{{\ttfamily
  2307.01852}}].

\bibitem{Appelquist:2024koa}
T.~Appelquist, J.~Ingoldby and M.~Piai, \emph{{Dilaton forbidden dark matter}},
  \href{https://doi.org/10.1103/PhysRevD.110.035013}{\emph{Phys. Rev. D}
  {\bfseries 110} (2024) 035013}
  [\href{https://arxiv.org/abs/2404.07601}{{\ttfamily 2404.07601}}].

\bibitem{Bruggisser:2022ofg}
S.~Bruggisser, B.~von Harling, O.~Matsedonskyi and G.~Servant, \emph{{Dilaton
  at the LHC: complementary probe of composite Higgs}},
  \href{https://doi.org/10.1007/JHEP05(2023)080}{\emph{JHEP} {\bfseries 05}
  (2023) 080} [\href{https://arxiv.org/abs/2212.00056}{{\ttfamily
  2212.00056}}].

\bibitem{Alonso:2025ksv}
R.~Alonso, S.~Chattopadhyay and J.~Ingoldby, \emph{{The Potential of HEFT and
  the scale of New Physics}},
  \href{https://arxiv.org/abs/2512.13612}{{\ttfamily 2512.13612}}.

\bibitem{Wu:2025hfp}
J.E.~Wu and Q.S.~Yan, \emph{{Confront a dilaton model with the LHC
  measurements}},  \href{https://arxiv.org/abs/2501.07820}{{\ttfamily
  2501.07820}}.

\bibitem{Zwicky:2023krx}
R.~Zwicky, \emph{{QCD with an infrared fixed point and a dilaton}},
  \href{https://doi.org/10.1103/PhysRevD.110.014048}{\emph{Phys. Rev. D}
  {\bfseries 110} (2024) 014048}
  [\href{https://arxiv.org/abs/2312.13761}{{\ttfamily 2312.13761}}].

\bibitem{Faedo:2024zib}
A.F.~Faedo, C.~Hoyos, M.~Piai, R.~Rodgers and J.G.~Subils, \emph{{Light
  holographic dilatons near critical points}},
  \href{https://doi.org/10.1103/PhysRevD.110.126017}{\emph{Phys. Rev. D}
  {\bfseries 110} (2024) 126017}
  [\href{https://arxiv.org/abs/2406.04974}{{\ttfamily 2406.04974}}].

\bibitem{Elander:2025fpk}
D.~Elander, A.F.~Faedo, M.~Piai, R.~Rodgers and J.G.~Subils, \emph{{Light
  dilaton near critical points in top-down holography}},
  \href{https://arxiv.org/abs/2502.19226}{{\ttfamily 2502.19226}}.

\bibitem{Cresswell-Hogg:2025kvr}
C.~Cresswell-Hogg, D.F.~Litim and R.~Zwicky, \emph{{Dilaton Physics from
  Asymptotic Freedom}},  \href{https://arxiv.org/abs/2502.00107}{{\ttfamily
  2502.00107}}.

\bibitem{Stegeman:2025sca}
R.~Stegeman and R.~Zwicky, \emph{{Gravitational $ D$-Form Factor: The
  $\sigma$-Meson as a Dilaton confronted with Lattice Data}},
  \href{https://arxiv.org/abs/2508.18537}{{\ttfamily 2508.18537}}.

\bibitem{Stegeman:2025tdl}
R.~Stegeman and R.~Zwicky, \emph{{Gluon Gravitational $ D$-Form Factor: The
  $\sigma$-Meson as a Dilaton confronted with Lattice Data II}},
  \href{https://arxiv.org/abs/2512.12315}{{\ttfamily 2512.12315}}.

\bibitem{SU(3)-data_release}
G.T.~Fleming and et~al. (Lattice Strong Dynamics~(LSD)), ``Dataset for ”light
  scalar meson and decay constant in su(3) gauge theory with eight dynamical
  flavors”.'' 2023.
\newblock \url{https://dx.doi.org/10.5281/zenodo.8007955}.

\bibitem{SU(2)-1-data_release}
A.~Athenodorou, E.~Bennett, G.~Bergner and B.~Lucini, ``Investigating the
  conformal behaviour of su(2) with one adjoint dirac flavor — data
  release.'' 2021.
\newblock \url{https://dx.doi.org/10.5281/zenodo.5139618}.

\bibitem{SU(2)-2-data_release}
A.~Athenodorou, E.~Bennett, G.~Bergner, J.~Lenz and B.~Lucini, ``Su(2) gauge
  theory with one and two adjoint fermions towards the continuum limit — data
  release.'' 2024.
\newblock \url{https://dx.doi.org/10.5281/zenodo.13128504}.

\bibitem{Cohen:1988sq}
A.G.~Cohen and H.~Georgi, \emph{{Walking Beyond the Rainbow}},
  \href{https://doi.org/10.1016/0550-3213(89)90109-0}{\emph{Nucl. Phys. B}
  {\bfseries 314} (1989) 7}.

\bibitem{Ryttov:2017kmx}
T.A.~Ryttov and R.~Shrock, \emph{{Higher-order scheme-independent series
  expansions of $\gamma_{\bar\psi\psi,IR}$ and $\beta'_{IR}$ in conformal field
  theories}}, \href{https://doi.org/10.1103/PhysRevD.95.105004}{\emph{Phys.
  Rev. D} {\bfseries 95} (2017) 105004}
  [\href{https://arxiv.org/abs/1703.08558}{{\ttfamily 1703.08558}}].

\bibitem{DelDebbio:2007wk}
L.~Del~Debbio, B.~Lucini, A.~Patella and C.~Pica, \emph{{Quenched mesonic
  spectrum at large N}},
  \href{https://doi.org/10.1088/1126-6708/2008/03/062}{\emph{JHEP} {\bfseries
  03} (2008) 062} [\href{https://arxiv.org/abs/0712.3036}{{\ttfamily
  0712.3036}}].

\bibitem{data_release}
T.~Appelquist, J.~Ingoldby and M.~Piai, ``Dilaton effective field theory across
  the conformal edge --- data release.'' 2025.
\newblock \url{https://doi.org/10.5281/zenodo.17966696}.

\end{thebibliography}\endgroup
\end{document}